\documentclass[twocolumn,prl,aps,showpacs]{revtex4}
\usepackage{graphicx}

\begin{document}

\title{\bf  Spectrum of the Vortex Bound States of the Dirac and Schrodinger Hamiltonian in the presence of Superconducting Gaps}

\author{Chi-Ho Cheng} \email{phcch@cc.ncue.edu.tw}
\affiliation{Department of Physics, National Changhua University
of Education, Taiwan}

\date{\today}

\begin{abstract}
We investigate the vortex bound states both Schrodinger and Dirac
Hamiltonian with the s-wave superconducting pairing gap by solving
the mean-field Bogoliubov-de-Gennes equations. The exact vortex
bound states spectrum is numerically determined by the integration
method, and also accompanied by the quasi-classical analysis. It
is found that the bound state energies is proportional to the
vortex angular momentum when the chemical potential is large
enough. By applying the external magnetic field, the vortex bound
state energies of the Dirac Hamiltonian are almost unchanged;
whereas the energy shift of the Schrodinger Hamiltonian is
proportional to the magnetic field. These qualitative differences
may serve as an indirect evidence of the existence of Majorana
fermions in which the zero mode exists in the case of the Dirac
Hamiltonian only.

\end{abstract}


\maketitle

\section{I. Introduction}
The zero mode (Majorana fermion excitation) attracts lots of
investigation due to its non-Abelian statistics
\cite{moore91,read,ivanov} and possible application to the fault
tolerant quantum processing \cite{kitaev,nayak}. Possible
candidates to support Majorana fermions are $p+ip$ superconductors
\cite{read,stern,stone}, $p+ip$ superfluids in cold atoms
\cite{gurarie,cheng,tewari}. With the proximity effect between the
$s$-wave superconductor and the strong topological insulator
surface \cite{fu07,moore07,fu07b,roy} , chiral Majorana fermions
could be created as edge states \cite{fu08}. Besides that,
Majorana fermions could also be realized in semiconductor with
spin-orbit coupling \cite{sau,alicea,potter}.

To verify the existence of Majorana fermion, Law {\it et al.}
proposed the tunneling experiments to probe the chiral Majorana
fermion at the interface between a superconductor and the surface
of a topological insulator, in which the Majorana fermions induce
resonant Andreev reflection \cite{law}.

In this paper, we investigate and compare the vortex bound states
of both the Dirac and Schrodinger Hamiltonian with the s-wave
superconducting gap by solving the mean-field Bogoliubov-de-Gennes
equations. The bound states are numerically solved by the
integration method as well as the quasi-classical analysis. Their
differences could be served as an indirect verification of the
existence of the Majorana fermion at the surface of the
topological insulator with an induced superconducting gap due to
the proximity effect.

\section{II. Formulation of vortex bound states}

Our formulation is based on the mean-field Bogoliubov-de-Gennes
(BdG) equations for the quasiparticle. In terms of Nambu indices,
the mean-field Hamiltonian becomes
\begin{eqnarray} \label{ham}
H &=& \frac{1}{2} \int d^2x \left(%
\begin{array}{cc}
  \Psi^\dagger & \Phi^\dagger \\
\end{array}%
\right)
\left(%
\begin{array}{cc}
  H_0^{(D,S)} & i\sigma_y\Delta(\vec r) \\
  -i\sigma_y\Delta^*(\vec r) & -H_0^{(D,S) *}  \\
\end{array}%
\right)
\left(%
\begin{array}{c}
  \Psi \\
  \Phi \\
\end{array}%
\right) \nonumber \\
\end{eqnarray}
where $\Psi = ( c_\uparrow , c_\downarrow )^T$ and $\Phi = (
c_\uparrow^\dagger , c_\downarrow^\dagger )^T$. $H_0^{(D,S)}$ in
this paper could be the Dirac Hamiltonian, $H_0^{(D)} = v_{\rm F}
\vec{p} \cdot \vec{\sigma} -\mu - h \sigma_z$, and the Schrodinger
Hamiltonian, $H_0^{(S)} = \frac{p^2}{2m} -\mu -h\sigma_z$. $v_{\rm
F}$ denotes the Fermi velocity, $\mu$ is the chemical potential,
$h$ represents the Zeeman coupling due to the external magnetic
field acting along the $z$-direction, and $\Delta(\vec r)$ is the
superconducting gap function.

The quasi-particle states of the above Hamiltonian in
Eq.(\ref{ham}) are given by
\begin{eqnarray} \label{quasi}
\left(%
\begin{array}{cc}
  H_0^{(D,S)} & i\sigma_y\Delta(\vec r) \\
  -i\sigma_y\Delta^*(\vec r) & -H_0^{(D,S) *}  \\
\end{array}%
\right) \left(%
\begin{array}{c}
  \textbf{u}_n(\vec r) \\
  \textbf{v}_n(\vec r) \\
\end{array}%
\right) = E_n \left(%
\begin{array}{c}
  \textbf{u}_n(\vec r) \\
  \textbf{v}_n(\vec r) \\
\end{array}%
\right)
\end{eqnarray}

Considering an isolated vortex carrying one flux quantum, {\it
i.e.} $\Delta(\vec r) = \Delta(r) {\rm e}^{i\phi}$, with the
eigenvectors of the form, $\textbf{u}_n(\vec r) = \left(
  u_{\uparrow,n}(r){\rm e}^{in\phi} ,  u_{\downarrow,n}(r){\rm
e}^{i(n+1)\phi} \right)$ and $\textbf{v}_n(\vec r) = \left(
  v_{\uparrow,n}(r){\rm e}^{in\phi} ,  v_{\downarrow,n}(r){\rm
e}^{i(n-1)\phi} \right)$. The eigenproblem for the Dirac
Hamiltonian in radial coordinate becomes
\begin{eqnarray} \label{bdg-dirac}
\left(%
\begin{array}{cc}
  K_n^{(D)} & i\sigma_y\Delta(r) \\
  -i\sigma_y\Delta(r) & -K_{n-1}^{(D) \dagger}  \\
\end{array}%
\right) \left(%
\begin{array}{c}
 \widetilde{u}_n(r) \\
 \widetilde{v}_n(r) \\
\end{array}%
\right) = E_n \left(%
\begin{array}{c}
  \widetilde{u}_n(r) \\
  \widetilde{v}_n(r) \\
\end{array}%
\right)
\end{eqnarray}
where
\begin{eqnarray}
K_n^{(D)} = \left(%
\begin{array}{cc}
  -\mu-h & -i \hbar v_{\rm F}(\frac{\partial}{\partial r}+\frac{n+1}{r}) \\
  -i \hbar v_{\rm F}(\frac{\partial}{\partial r}-\frac{n}{r}) & -\mu+h \\
\end{array}%
\right)
\end{eqnarray}
and ${\widetilde u}_n(r) = \left( u_{\uparrow,n}(r) ,
u_{\downarrow,n}(r) \right)^T$, $\widetilde{v}_n(r) = \left(%
v_{\uparrow,n}(r) , v_{\downarrow,n}(r) \right)^T$.

For the Schrodinger Hamiltonian, the eigenproblem can be further
block-diagonalized into
\begin{eqnarray} \label{bdg-sch}
(K_n^{(S)}  + \sigma_x \Delta(r) - h)
\left(%
\begin{array}{c}
 u_{\uparrow,n}(r) \\
 v_{\downarrow,n}(r) \\
\end{array}%
\right) = E_n \left(%
\begin{array}{c}
  u_{\uparrow,n}(r) \\
  v_{\downarrow,n}(r) \\
\end{array}%
\right) \nonumber \\
\end{eqnarray}
\begin{eqnarray} \label{bdg-sch-2}
(K_{n+1}^{(S)}  - \sigma_x \Delta(r) + h)
\left(%
\begin{array}{c}
 u_{\downarrow,n}(r) \\
 v_{\uparrow,n}(r) \\
\end{array}%
\right) = E_n \left(%
\begin{array}{c}
  u_{\downarrow,n}(r) \\
  v_{\uparrow,n}(r) \\
\end{array}%
\right) \nonumber \\
\end{eqnarray}
with
\begin{widetext}
\begin{eqnarray}
K_n^{(S)} =   \left(%
\begin{array}{cc}
  -\frac{\hbar^2}{2m}(\frac{\partial^2}{\partial r^2} + \frac{1}{r}\frac{\partial}{\partial r} - \frac{n^2}{r^2}) -\mu & 0 \\
  0 & \frac{\hbar^2}{2m}(\frac{\partial^2}{\partial r^2} + \frac{1}{r}\frac{\partial}{\partial r} - \frac{(n-1)^2}{r^2}) +\mu \\
\end{array}%
\right)
\end{eqnarray}
\end{widetext}
Eqs.(\ref{bdg-sch}) and (\ref{bdg-sch-2}) are not independent to
each other. In fact, their complete sets of eigenvectors are the
same.

The size of the vortex core is characterized by the coherent
length $\xi$. For the Dirac Hamiltonian, it should be of the order
of $\hbar v_{\rm F}/\Delta_0$, the unique length scale of the
system, where $\Delta_0$ is the superconducting gap far away from
the vortex core. However, for the Schrodinger Hamiltonian, there
are two length scales $k_{\rm F}^{-1}$ and $\hbar v_{\rm
F}/\Delta_0$. $\xi$ is of the order of $\hbar v_{\rm F}/\Delta_0$
when $\mu/\Delta_0 \gg 1$. As $\mu/\Delta_0$ becomes smaller,
$\xi$ is going to shrink to another length scale, $k_{\rm F}^{-1}$
\cite{kramer, gygi, sensarma}. When $\mu/\Delta_0$ turns to be
negative, a new length scale due to the bosonic molecules emerges.

We adopt $\xi=\hbar v_{\rm F}/\Delta_0$ for the Dirac Hamiltonian,
and $\xi=\hbar c/\Delta_0$ for the Schrodinger Hamiltonian in
which $c$ is a constant independent of the chemical potential
$\mu$. Further we assume the form of the gap amplitude
\begin{eqnarray} \label{gap-amplitude}
\Delta(r) = \Delta_0 \tanh(\frac{r}{\xi})
\end{eqnarray}

The BdG equations, Eqs.(\ref{bdg-dirac}) and (\ref{bdg-sch}), are
solved by the integration method. Imposing that the wavefunction
aymptotically approaches to zero far away from the vortex core,
the eigenenergies can be obtained by integrating the differential
equation from the infinity to the vortex center. The advantage of
the integration method is that we can solve for the numerically
exact (that is, system size in the thermodynamical limit) bound
state energies once the form of the gap amplitude as in
Eq.(\ref{gap-amplitude}) is given.

The chemical potential $\mu$ in the BdG equations for the
Schrodinger Hamiltonian can be positive or negative, arbitrary
small or large. The condition of the quasi-classical
approximation, $k_{\rm F} \xi \gg 1$ (equivalent to $\mu/\Delta_0
\gg 1$), may not be valid in the whole range of the chemical
potential. Eq.(\ref{bdg-sch}) is solved directly by the
integration method. We do not assume the Andreev (quasi-classical)
approximation \cite{clinton,kramer,gygi}.

\section{III. Numerical Results}


We first consider the case without external magnetic field. The
vortex bound state energies of the Dirac Hamiltonian are solved
for different chemical potential.

\begin{figure}[tbh]
\begin{center}
\includegraphics[width=3in]{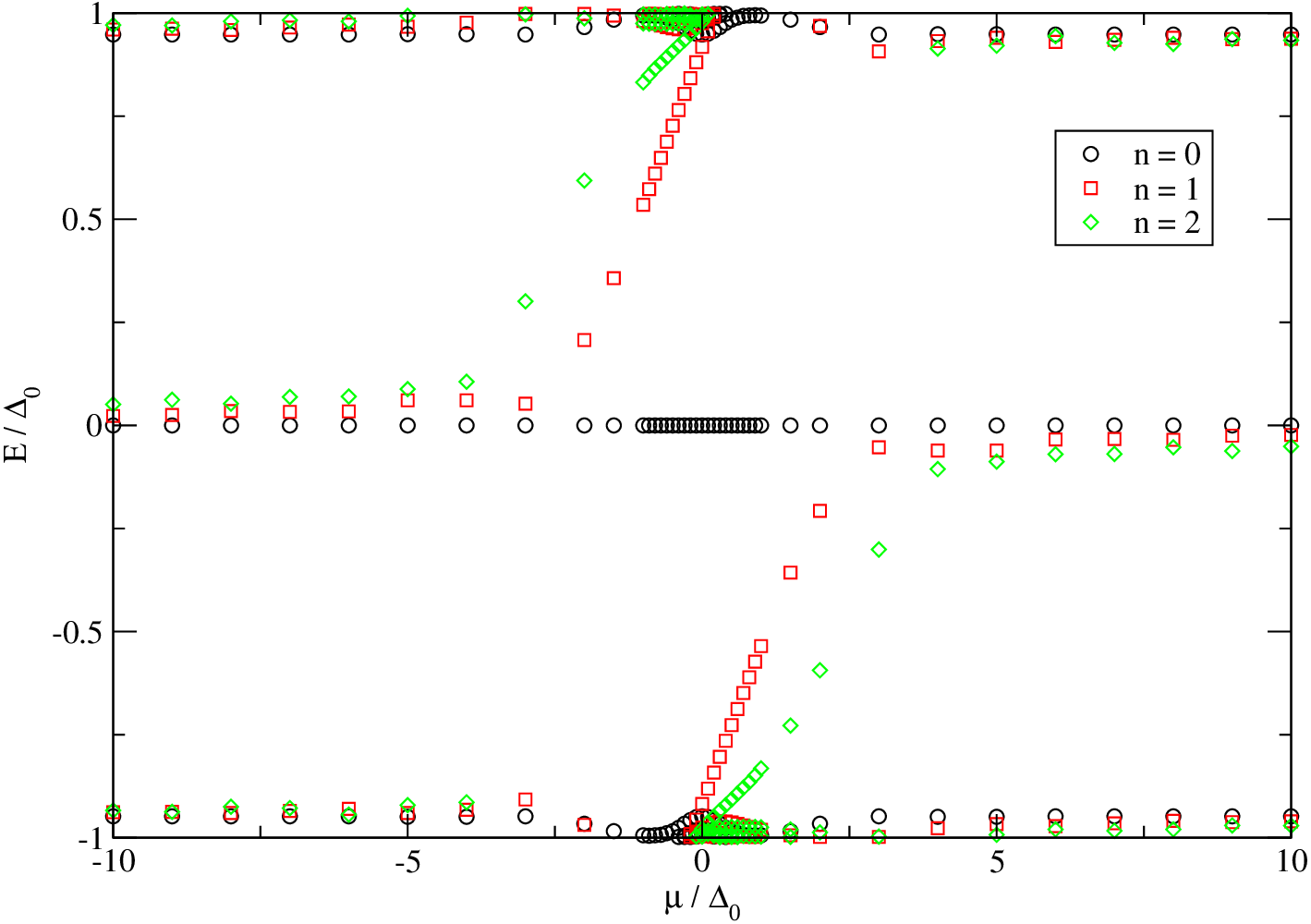}
\end{center}
\vspace{-5pt}
 \caption{Vortex bound state energies of the Dirac Hamiltonian as a function of chemical potential $\mu$ in the absence of
 external magnetic field, $h=0$, for different angular momentum label by quantum number $n$. $E_{-n}=-E_n$.}
 \label{dirac-erg-mu.eps}
 \vspace{15pt}
\end{figure}

The vortex bound state energies of the Dirac Hamiltonian as
function of chemical potential $\mu$ is shown in
Fig.\ref{dirac-erg-mu.eps}. We do not show the negative $n$
because $E_{-n} = - E_n$ which is the consequence of time-reversal
symmetry. That is, $E \rightarrow -E$ when $n \rightarrow -n$
(reverse the angular momentum). Besides that, $E \rightarrow -E$
when $\mu \rightarrow -\mu$, which is the particle-hole symmetry
particularly for the Dirac Hamiltonian.

For $n=0$, the zero mode corresponds to the Majorana fermions. For
non-zero angular momentum, $n > 0$, $|\mu| \gg \Delta_0$, we found
the mid-gap state near zero energy. As $\mu$ approaches to zero,
the mid-gap state gradually merges to the continuum of energy $E =
\Delta_0$. Especially, near zero chemical potential, $\mu\simeq
0$, all the vortex bound states except the Majorana fermions
state, are located near the edge.

\begin{figure}[tbh]
\begin{center}
\includegraphics[width=3in]{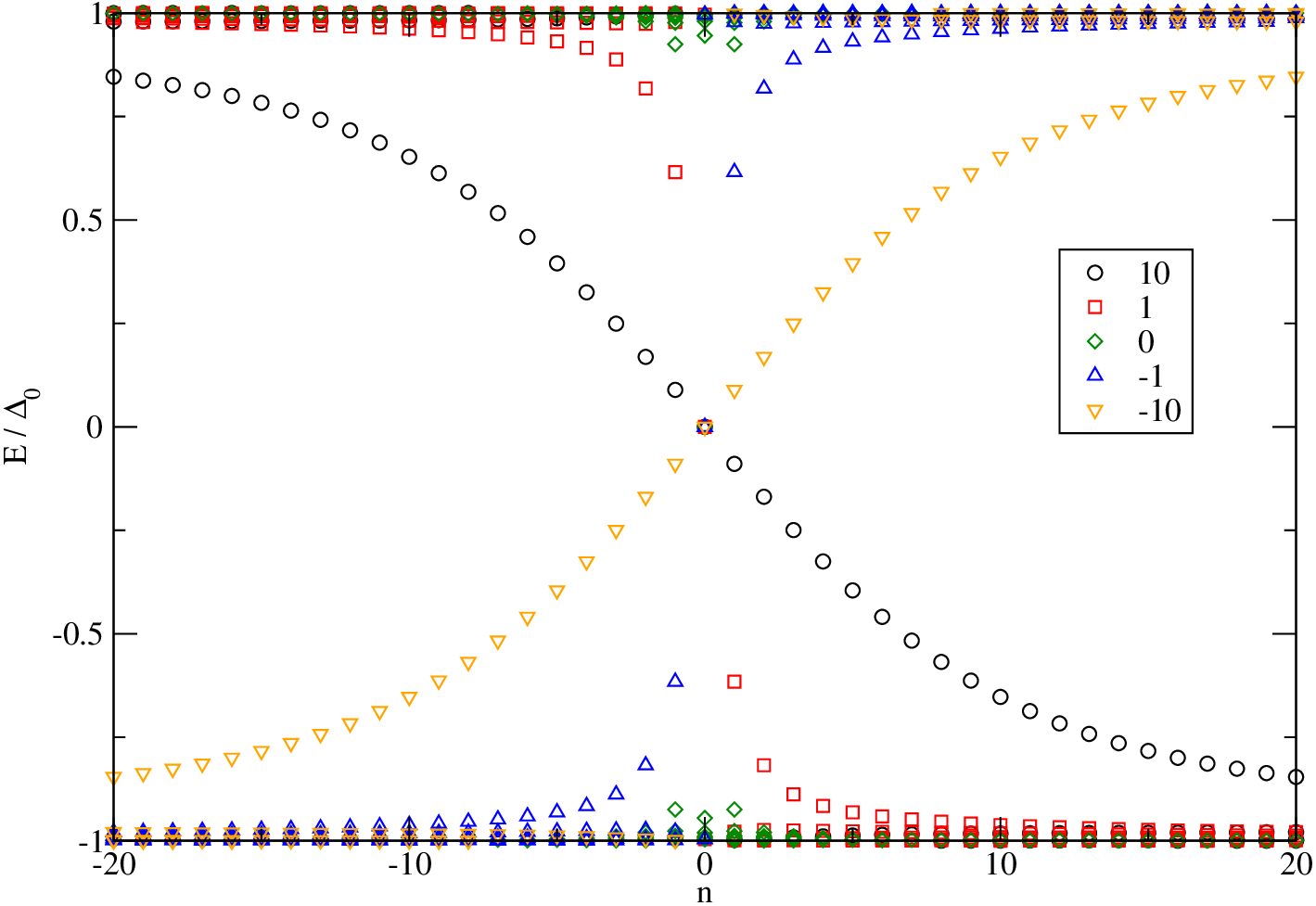}
\end{center}
\vspace{-5pt}
 \caption{Vortex bound state energies of the Dirac Hamiltonian as a function of angular momentum quantum numbers $n$ in the absence of
 external magnetic field, $h=0$, for $\mu/\Delta_0 = 10, 1, 0, -1, -10$.}
 \label{dirac-erg-n.eps}
 \vspace{15pt}
\end{figure}

We also plot the bound state energy as a function of the angular
momentum quantum number $n$, as shown in
Fig.\ref{dirac-erg-n.eps}. There is a simple linear relationship
$E \propto n$ for small $|n|$ (up to $n = 10$) when $\mu \gg
\Delta_0$. This linear relationship could be explained by
quasi-classical analysis in the next section. As $\mu \simeq 0$,
the bound state energy except the zero mode merges into the
continuum $\Delta_0$. The prediction of the bound state energies
by linearizing the vortex amplitude \cite{lu,herbut} is
proportional to $\sqrt{N}$ ($N$ is an integer quantum number),
which is inconsistent with our calculation.

\begin{figure}[tbh]
\begin{center}
\includegraphics[width=3in]{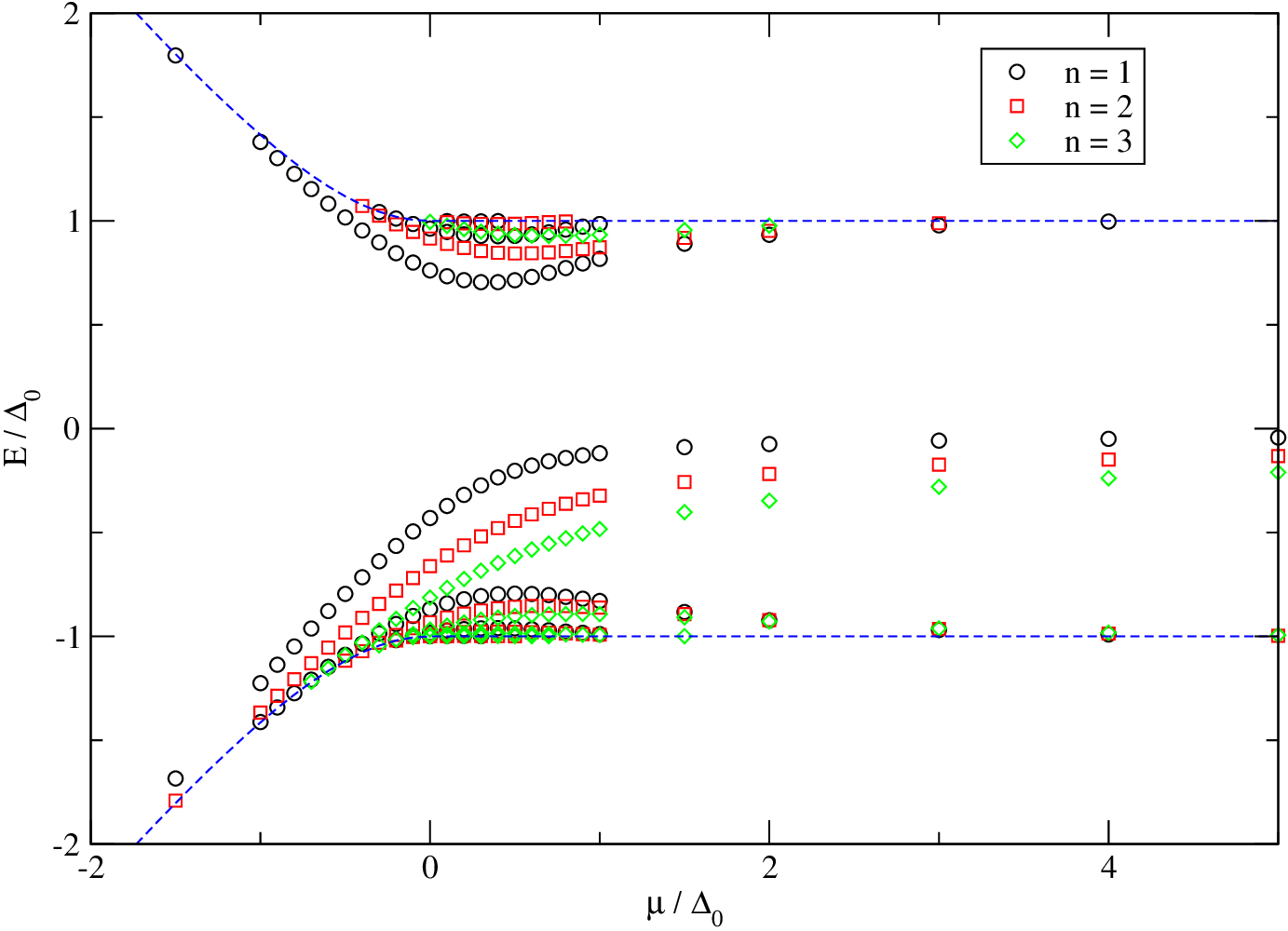}
\end{center}
\vspace{-5pt}
 \caption{Vortex bound state energies of the Schrodinger Hamiltonian as a function of chemical potential $\mu$ in the absence of
 external magnetic field, $h=0$, for different angular momentum label by quantum number $n$. $m c^2/\Delta_0 = 10$. $E_{-n}=-E_{n-1}$.
 The dashed lines  represent the continuum threshold.}
 \label{sch-erg-mu.eps}
 \vspace{15pt}
\end{figure}
\begin{figure}[tbh]
\begin{center}
\includegraphics[width=3in]{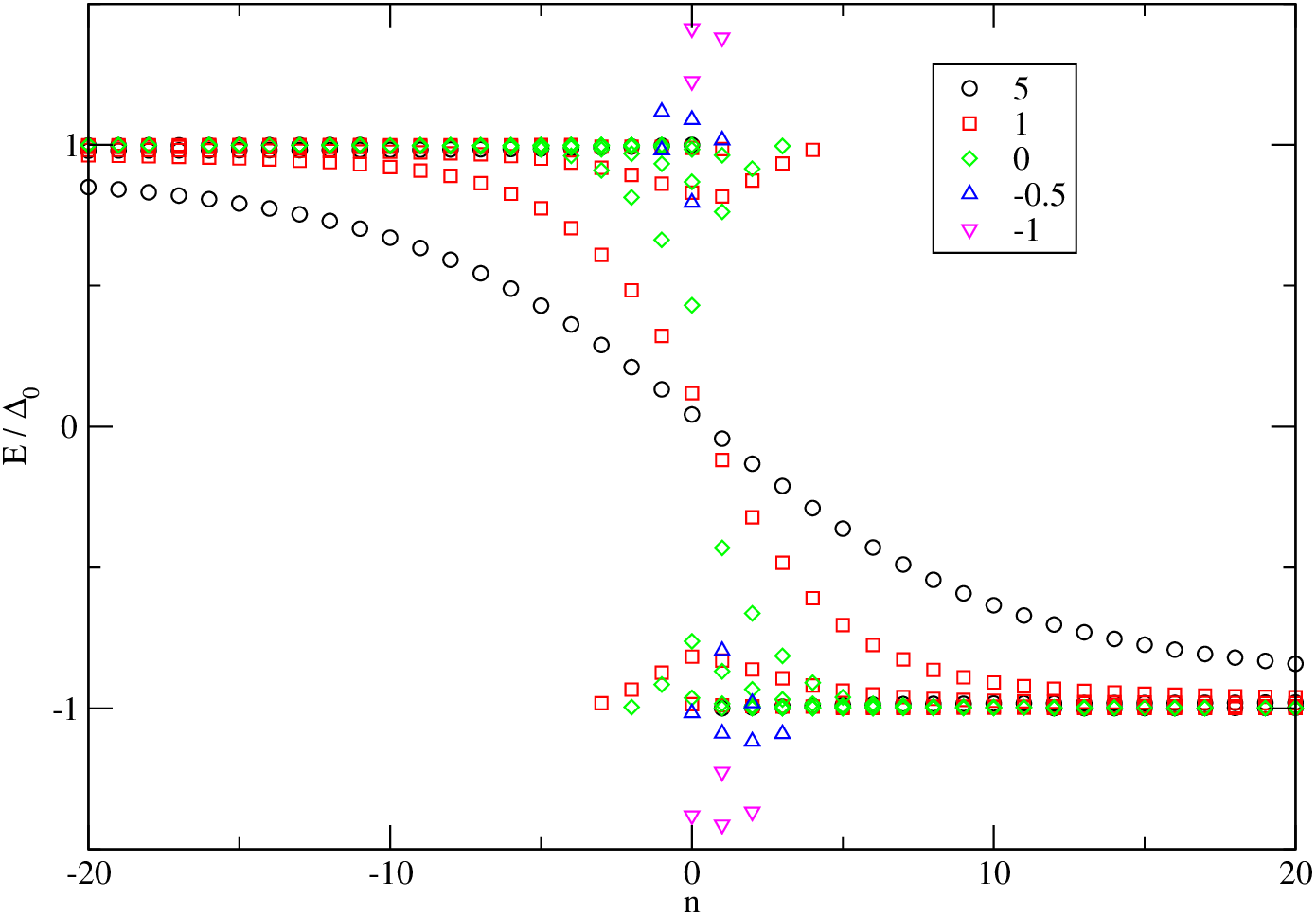}
\end{center}
\vspace{-5pt}
 \caption{Vortex bound state energies of the Schrodinger Hamiltonian as a function of angular momentum quantum numbers $n$ in the absence of
 external magnetic field, $h=0$, for $\mu/\Delta_0 = 5, 1, 0, -0.5, -1$ when $m c^2/\Delta_0 = 10$.}
 \label{sch-erg-n.eps}
 \vspace{15pt}
\end{figure}
\begin{figure}[tbh]
\begin{center}
\includegraphics[width=3in]{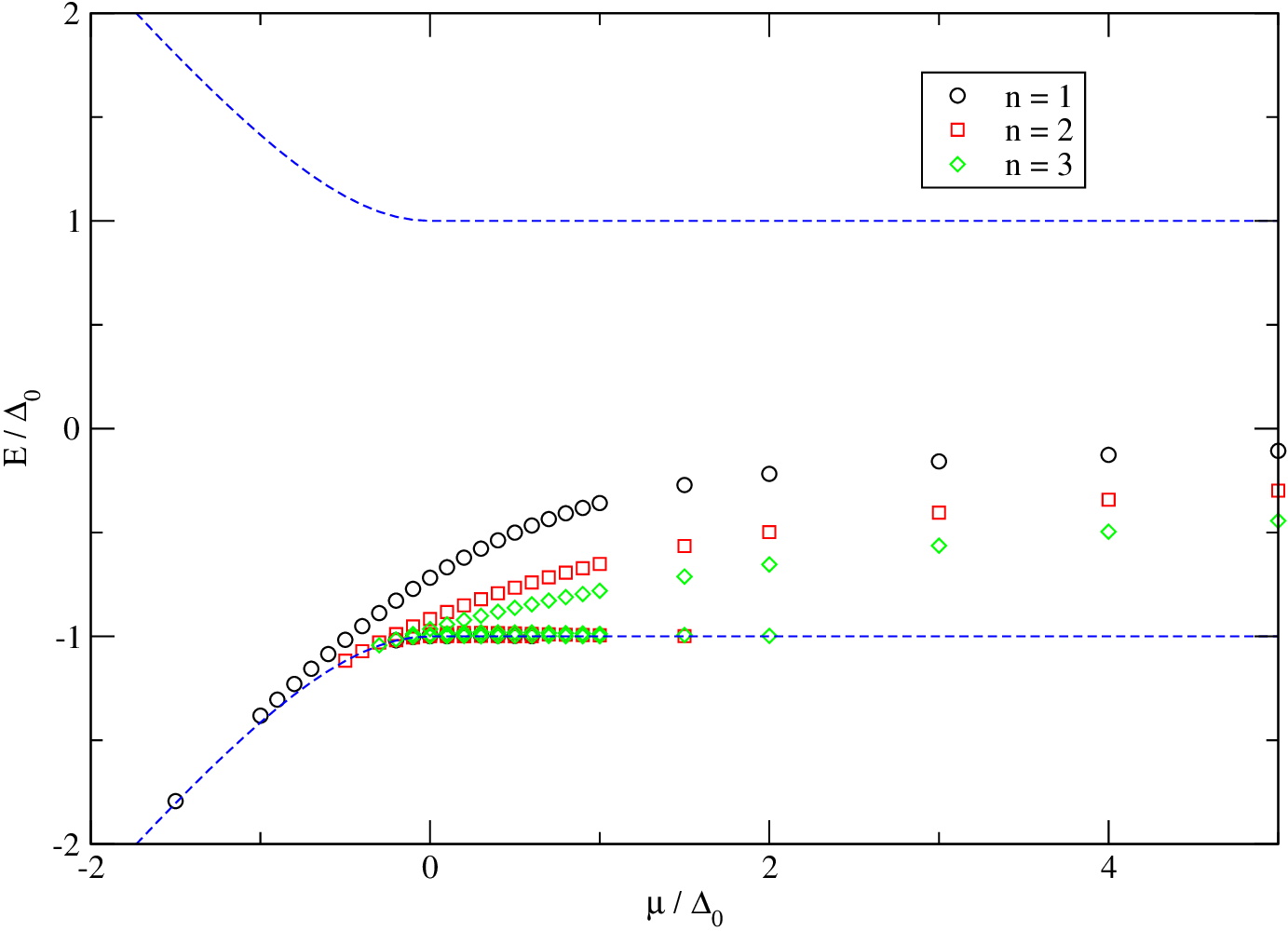}
\end{center}
\vspace{-5pt}
 \caption{Vortex bound state energies of the Schrodinger Hamiltonian as a function of chemical potential $\mu$ in the absence of
 external magnetic field, $h=0$, for different angular momentum label by quantum number $n$. $m c^2/\Delta_0 = 1$. $E_{-n}=-E_{n-1}$.
 The dashed lines  represent the continuum threshold. }
 \label{sch-erg-mu_c1.eps}
 \vspace{15pt}
\end{figure}
\begin{figure}[tbh]
\begin{center}
\includegraphics[width=3in]{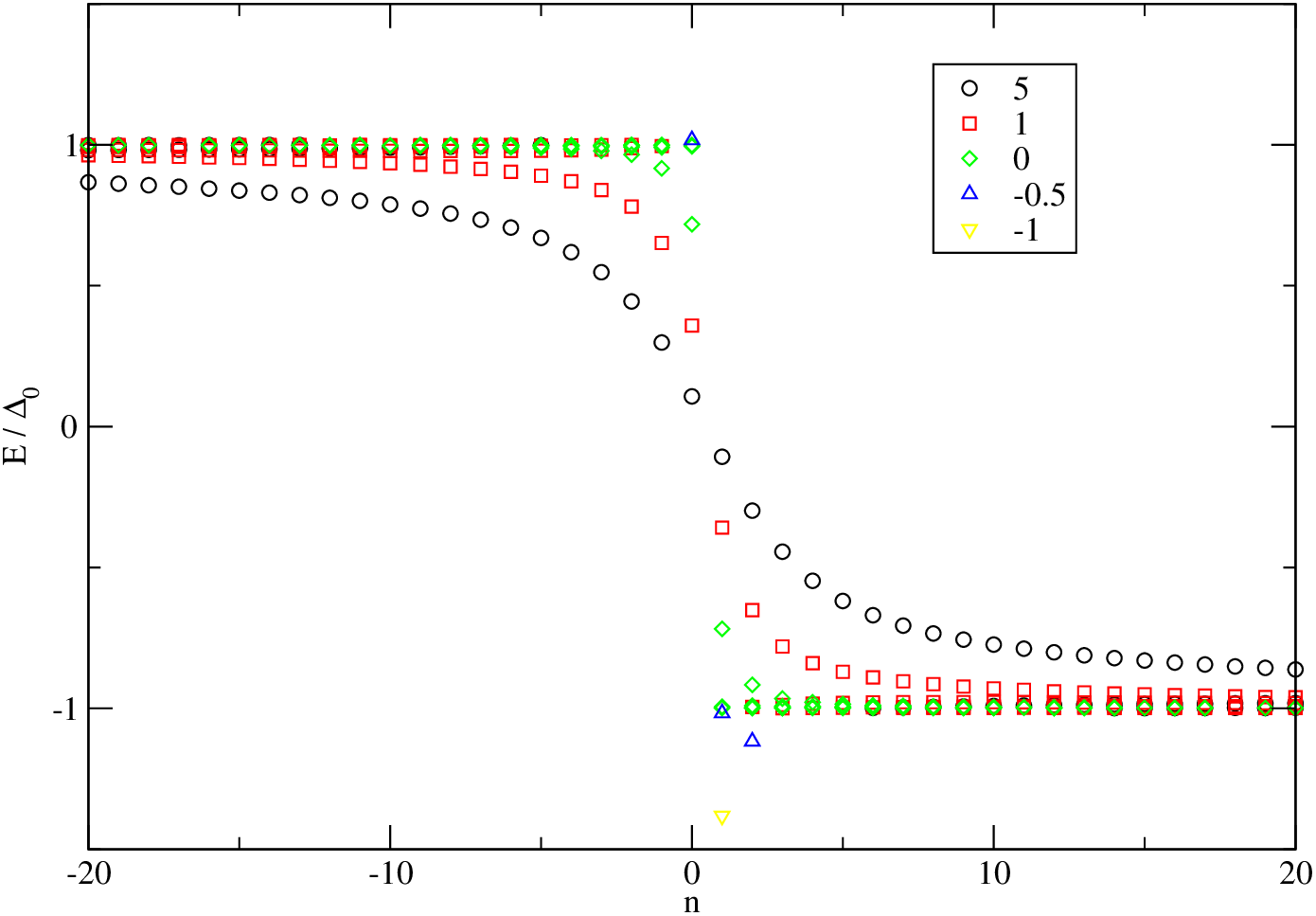}
\end{center}
\vspace{-5pt}
 \caption{Vortex bound state energies of the Schrodinger Hamiltonian as a function of angular momentum quantum numbers $n$ in the absence of
 external magnetic field, $h=0$, for $\mu/\Delta_0 = 5, 1, 0, -0.5, -1$ when $m c^2/\Delta_0 = 1$.}
 \label{sch-erg-n_c1.eps}
 \vspace{15pt}
\end{figure}

We also solve for the bound state energies of the Schrodinger
Hamiltonian for comparison. Fig.\ref{sch-erg-mu.eps} shows the
bound state energies as a function of $\mu$, with $m c^2 /\Delta_0
= 10$. Due to the time-reversal symmetry, $E_{-n} = - E_{n-1}$.
Unlike the Dirac case, no particle-hole symmetry remains. For
highly positive chemical potential $\mu \gg \Delta_0$, the
qualitative behavior is the same as the Dirac case except that the
zero mode is missing. A simple linear relationship $E \propto n$
is also recovered, as shown in Fig.\ref{sch-erg-n.eps}. The
negative chemical potential, $\mu < 0$, can be realized in
superfluid Fermi gases near the BEC to BCS crossover
\cite{sensarma}. For $\mu < 0$, the continuum of the excitation
becomes $\sqrt{\mu^2+\Delta_0^2}$. The mid-gap state merges into
the continuum when the chemical potential is deeply negative, $\mu
\ll -\Delta_0$.  The qualitative behavior of the mid-gap states
remains the same for different $c$, which is also shown in
Figs.\ref{sch-erg-mu_c1.eps} and \ref{sch-erg-n_c1.eps} with
$mc^2/\Delta_0=1$.


Under the external magnetic field perpendicular to the surface,
the vortex bound states of the Dirac and Schrodinger Hamiltonian
behave very differently.

For the Dirac Hamiltonian, the continuum threshold of bound state
energy is $\Delta_0\sqrt{1-h^2/\Delta_0^2}$ when $0< h <
\mu/\sqrt{1+\Delta_0^2/\mu^2}$, and $\sqrt{\mu^2+\Delta_0^2}-h$
when $\mu/\sqrt{1+\Delta_0^2/\mu^2} < h <
\sqrt{\mu^2+\Delta_0^2}$, and the energy becomes fully continuum
when $h > \sqrt{\mu^2+\Delta_0^2}$.

\begin{figure}[tbh]
\begin{center}
\includegraphics[width=3in]{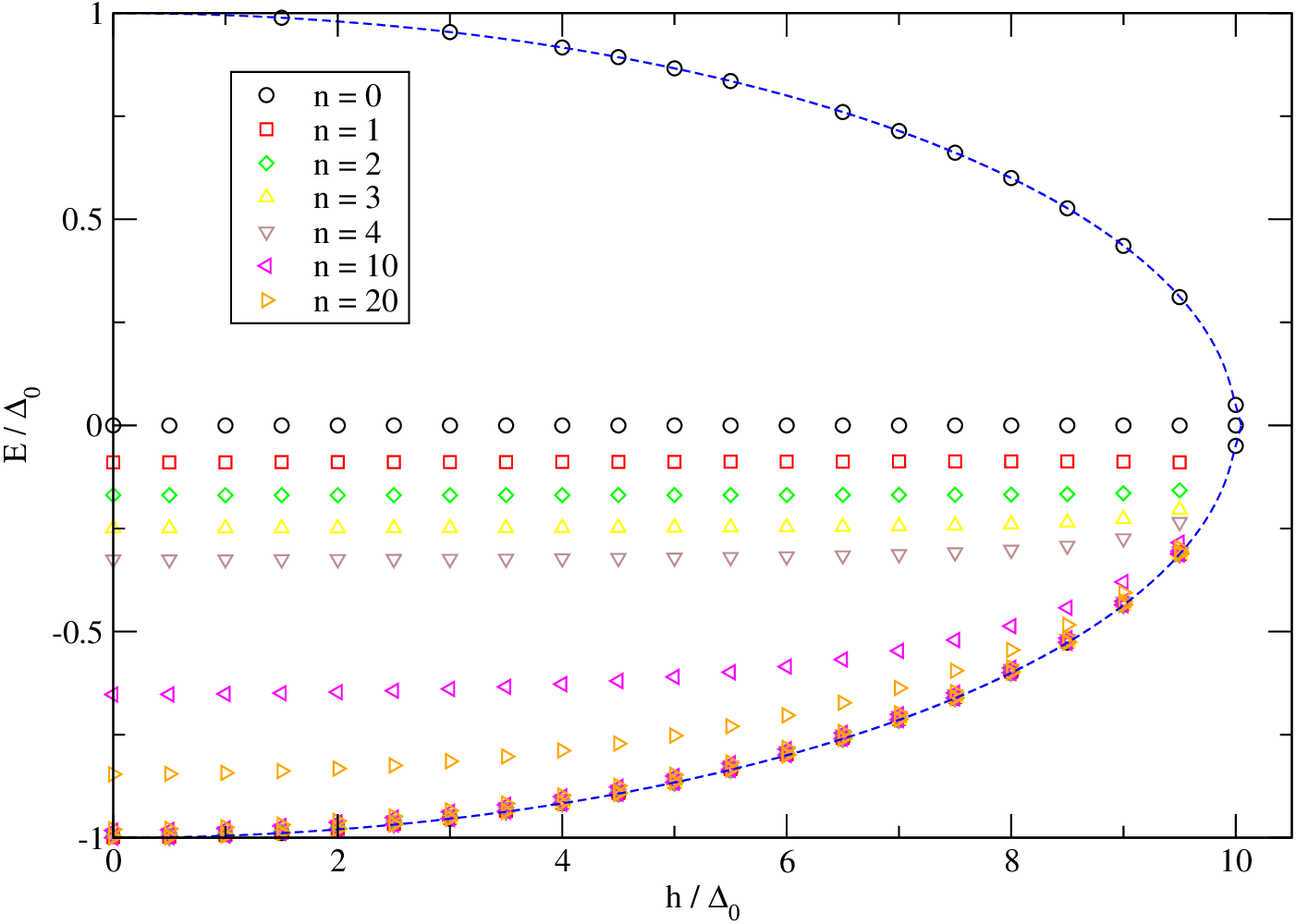}
\end{center}
\vspace{-5pt}
 \caption{Vortex bound state energies of the Dirac Hamiltonian as a function of external magnetic field $h$
 at $\mu/\Delta_0 =10$, for different $n$. The dashed lines  represent the continuum threshold.}
 \label{dirac-erg-h_mu10.eps}
 \vspace{15pt}
\end{figure}

In the limit of $|\mu| \gg \Delta_0$, as illustrated in
Fig.\ref{dirac-erg-h_mu10.eps} for $\mu/\Delta_0 = 10$, the vortex
bound states of different angular momenta are equally spaced. As
the external magnetic field increases, the vortex bound states
merge into the continuum. Notice that all the states are
insensitive under the external magnetic field.

\begin{figure}[tbh]
\begin{center}
\includegraphics[width=3in]{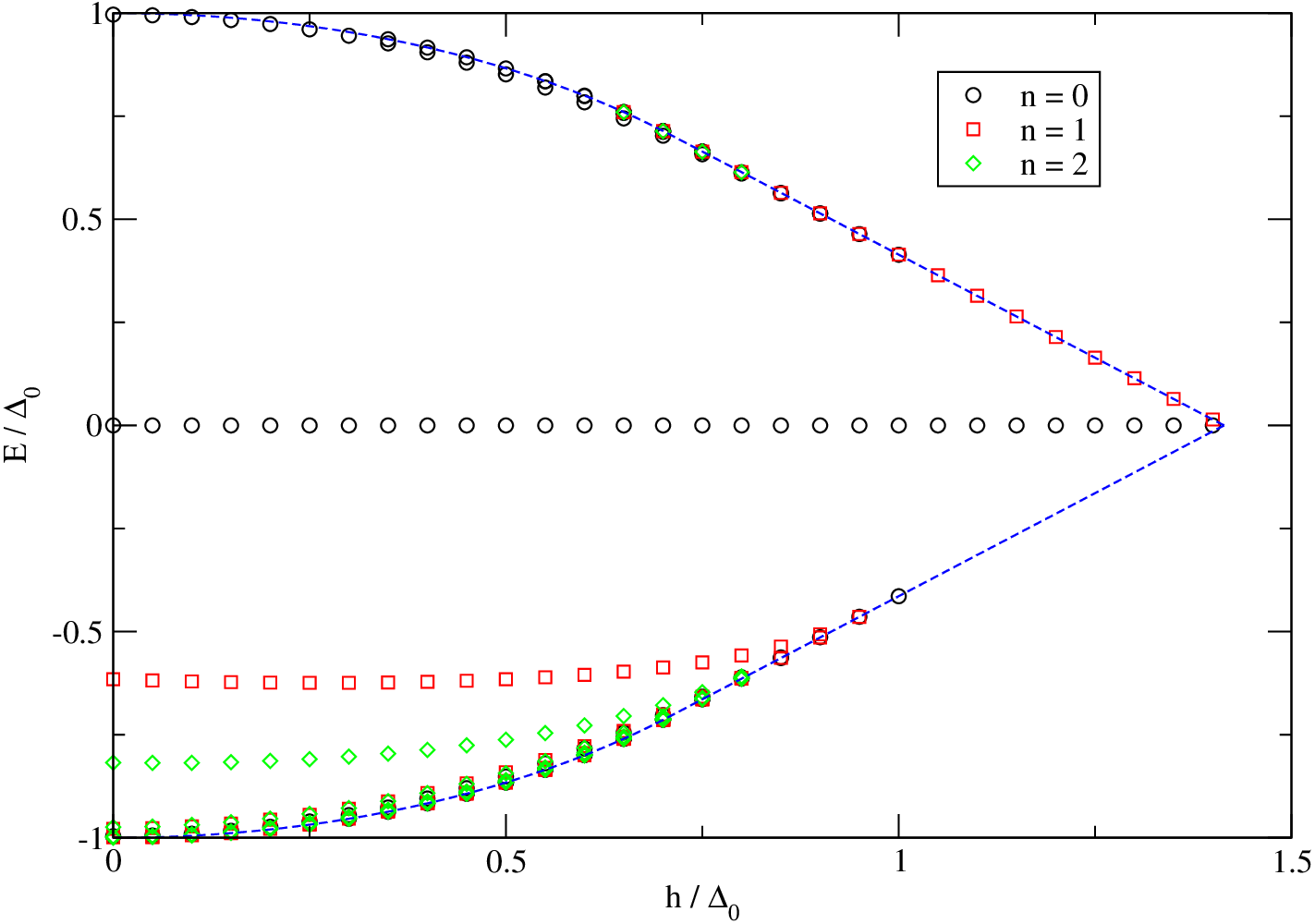}
\end{center}
\vspace{-5pt}
 \caption{Vortex bound state energies of the Dirac Hamiltonian as a function of external magnetic field $h$
 at $\mu/\Delta_0 =1$, for different $n$. The dashed lines  represent the continuum threshold.}
 \label{dirac-erg-h_mu1.eps}
 \vspace{15pt}
\end{figure}

\begin{figure}[tbh]
\begin{center}
\includegraphics[width=3in]{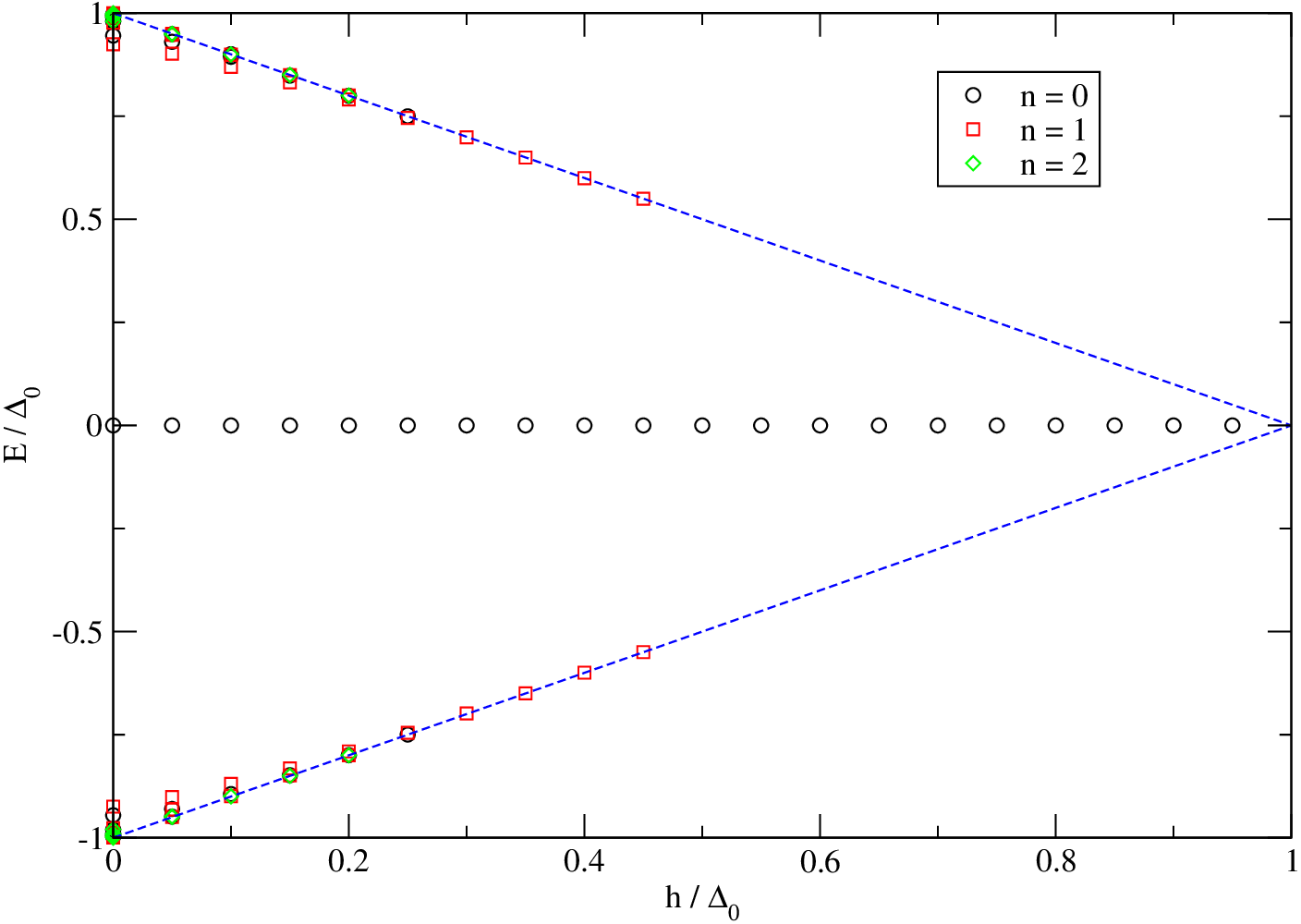}
\end{center}
\vspace{-5pt}
 \caption{Vortex bound state energies of the Dirac Hamiltonian as a function of external magnetic field $h$
 at $\mu/\Delta_0 =0$, for different $n$. The dashed lines  represent the continuum threshold.}
 \label{dirac-erg-h_mu0.eps}
 \vspace{15pt}
\end{figure}

As $|\mu|$ decreases, the vortex bound states of finite angular
momentum moves towards the continuum, as illustrated in
Fig.\ref{dirac-erg-h_mu1.eps} for $\mu/\Delta_0 = 1$. When $|\mu|
\ll \Delta_0$, the zero mode exists until it meets the continuum
when $h > \sqrt{\mu^2+\Delta_0^2}$. The remaining bound states are
located near the edge of continuum. The case that $\mu/\Delta_0 =
0$ is shown in Fig.\ref{dirac-erg-h_mu0.eps}.

For the Schrodinger Hamiltonian, it is easy to see from
Eqs.(\ref{bdg-sch})-(\ref{bdg-sch-2}) that the vortex bound state
energy shift due to the external magnetic field is $\pm h$. This
is the fundamental difference from the Dirac Hamiltonian.

\section{IV. Quasi-classical Analysis}
\subsection{A. Dirac Hamiltonian}

From Eq.(\ref{quasi}), the eigenvector of the zero energy state is
given by
\begin{eqnarray}
\left(%
\begin{array}{cc} \label{bdg}
  v_{\rm F}\vec{p}\cdot\vec{\sigma}-\mu -h\sigma_z & i\sigma_y\Delta \\
  -i\sigma_y\Delta^* & -v_{\rm F}\vec{p}\cdot\vec{\sigma}+\mu +h\sigma_z \\
\end{array}%
\right)
\left(%
\begin{array}{c}
  \textbf{u} \\
  \textbf{v} \\
\end{array}%
\right) = 0 \nonumber \\
\end{eqnarray}

Under the condition of the quasi-classical approximation, {\it
i.e.}, $k_{\rm F} \xi \gg 1$, the wavelength of the
quasi-particles is much smaller than the coherent length of the
vortex, their trajectories are almost straight lines. Following
the argument by Volovik \cite{volovik}, the zero energy state
corresponds to the trajectory that crosses the center of the
vortex, {\it i.e.}, zero impact parameter, $b = 0$. The
superconducting gap function
\begin{eqnarray} \label{deltas}
\Delta(s) = \Delta(|s|) {\rm sgn}(s)
\end{eqnarray}
where $s$ parameterizes the trajectory across the vortex center.

For $|h| < |\mu|$, by directly solving
Eqs.(\ref{bdg})-(\ref{deltas}), we have the zero energy state
\begin{eqnarray}
\left(%
\begin{array}{c}
  \textbf{u} \\
  \textbf{v} \\
\end{array}%
\right) = \frac{{\rm e}^{i q s} I\bigotimes{\rm e}^{\frac{i}{2} \sigma_z \theta}}{2 \sqrt{\mu(\mu-h)}}  \left(%
\begin{array}{c}
  (\mu - h) \\
  \sqrt{\mu^2 - h^2} \\
  i(\mu - h) \\
  -i \sqrt{\mu^2 - h^2} \\
\end{array}%
\right) \psi(s) \nonumber \\
\end{eqnarray}
where the plane wave momentum ${\vec q} =
\frac{\sqrt{\mu^2-h^2}}{\hbar v_{\rm F}}({\rm \hat e}_x \cos\theta
+ {\rm \hat e}_y \sin\theta)$, and
\begin{eqnarray} \label{psi}
\psi(s) = \exp{\left(-\frac{1}{\hbar v_{\rm F}}\int^s ds'
\Delta(|s'|) {\rm sgn}(s') \right)}
\end{eqnarray}

The non-zero angular momentum vortex bound state near the zero
energy state can be estimated by small impact parameter $b$,
$\Delta(r=\sqrt{s^2+b^2}){\rm e}^{i\phi} \simeq \Delta(|s|){\rm
sgn(s)} + i\Delta(|s|)b/s$. Hence we have the perturbation
\begin{eqnarray}
H' = \frac{i\Delta(|s|) b}{s} \left(%
\begin{array}{cc}
  0 & i\sigma_y  \\
  i\sigma_y & 0 \\
\end{array}%
\right)
\end{eqnarray}
The bound state energies up to first order correction becomes
\begin{eqnarray}
E = \langle H' \rangle = {\rm sgn}(\mu)  b
\frac{\int_{-\infty}^\infty ds \frac{\Delta(|s|)}{s} |\psi(s)|^2
}{ \int_{-\infty}^\infty ds
 |\psi(s)|^2}
\end{eqnarray}
and the angular momentum $L_z = -p b = -b|\mu|/ v_{\rm F}$ which
is $n \hbar$ by the semi-classical quantization condition,
\begin{eqnarray}
E = -n \hbar \omega_0
\end{eqnarray}
with
\begin{eqnarray} \label{w0}
\omega_0 = \frac{v_{\rm F}}{\mu} \frac{\int_{-\infty}^\infty ds
\frac{\Delta(|s|)}{s} |\psi(s)|^2 }{ \int_{-\infty}^\infty ds
 |\psi(s)|^2}
\end{eqnarray}
Notice that $\omega_0$ has no $h$ dependence.

On the other hand, when $|h| > |\mu|$, the zero energy state is a
decaying solution,
\begin{eqnarray}
\left(%
\begin{array}{c}
  \textbf{u} \\
  \textbf{v} \\
\end{array}%
\right) = \frac{{\rm e}^{- \kappa s} I\bigotimes{\rm e}^{\frac{i}{2} \sigma_z \theta}}{2 \sqrt{\mu(\mu-h)}}  \left(%
\begin{array}{c}
  h - \mu \\
  -i \sqrt{h^2 - \mu^2} \\
  i (h - \mu ) \\
   \sqrt{h^2 - \mu^2} \\
\end{array}%
\right) \psi(s) \nonumber \\
\end{eqnarray}
where $\kappa=\frac{\sqrt{h^2-\mu^2}}{\hbar v_{\rm F}}$. The
correction $\langle H' \rangle = 0$.  It is consistent with all
finite angular momentum bound states merging to zero bound state
at $h=\mu$.

\subsection{B. Schrodinger Hamiltonian}

For the case of the Schrodinger Hamiltonian with superconducting
gap, the eigenproblem for the quasi-particles is
\begin{eqnarray}
\left( \left( \frac{p^2}{2m} -\mu \right) \sigma_z + \left(%
\begin{array}{cc}
  0 & \Delta \\
  \Delta^* & 0 \\
\end{array}%
\right) - h \right)
\left(%
\begin{array}{c}
 u_{\uparrow} \\
 v_{\downarrow} \\
\end{array}%
\right) = E \left(%
\begin{array}{c}
  u_{\uparrow} \\
  v_{\downarrow} \\
\end{array}%
\right) \nonumber \\
\end{eqnarray}

In the quasi-classical approach, after the transformation
$(u_\uparrow, v_\downarrow) \rightarrow {\rm e}^{i {\vec q}\cdot
{\vec r}} (u_\uparrow, v_\downarrow)$, the eigenproblem becomes
\begin{widetext}
\begin{eqnarray}
\left( \left( {\vec v}_{\rm F}\cdot{\vec p} -\frac{\hbar^2}{2mr^2}\frac{\partial^2}{\partial \phi^2} \right) \sigma_z + \left(%
\begin{array}{cc}
  0 & \Delta \\
  \Delta^* & 0 \\
\end{array}%
\right) - h \right)
\left(%
\begin{array}{c}
 u_{\uparrow} \\
 v_{\downarrow} \\
\end{array}%
\right) = E \left(%
\begin{array}{c}
  u_{\uparrow} \\
  v_{\downarrow} \\
\end{array}%
\right) \\
\left( ({\vec v}_{\rm F}\cdot{\vec p} ) \sigma_z + \sigma_x
\Delta(s) \right)
\left(%
\begin{array}{c}
 u_{\uparrow} \\
 v_{\downarrow} \\
\end{array}%
\right) = \left( E +h +\frac{\hbar^2 \sigma_z}{2mr^2}\frac{\partial^2}{\partial \phi^2} - i\sigma_y \frac{i\Delta(|s|)b}{s}) \right)\left(%
\begin{array}{c}
  u_{\uparrow} \\
  v_{\downarrow} \\
\end{array}%
\right)
\end{eqnarray}
\end{widetext}
where ${\vec q}=\frac{\sqrt{2m \mu}}{\hbar}({\rm \hat e}_x
\cos\theta + {\rm \hat e}_y \sin\theta)$, and ${\vec v}_{\rm
F}=\hbar {\vec q}/m$, for $\mu > 0$.

We solve the left hand side, and treat the right hand side to the
first order perturbation \cite{degennes}. The bound state energy
is
\begin{eqnarray}
E = -(n-\frac{1}{2})\hbar \omega_0 -h
\end{eqnarray}
with $\omega_0$ defined in Eq.(\ref{w0}). $n$ is an integer. The
half integer $\frac{1}{2}$ is due to the contribution of the term
$-\frac{\hbar^2 \sigma_z}{2mr^2}\frac{\partial^2}{\partial
\phi^2}$ which is absent in the Dirac Hamiltonian. Because of the
extra half integer, there is no zero mode in the vortex bound
state of the Schrodinger Hamiltonian.

The bound state energy for $(u_\downarrow, v_\uparrow)$ can be
obtained from the particle-hole transformation, which gives
\begin{eqnarray}
E = (n-\frac{1}{2})\hbar \omega_0 +h
\end{eqnarray}

\section{V. Conclusion}

In conclusion, we solve numerically and perform the
quasi-classical analysis for the vortex bound states of both the
Dirac and Schrodinger Hamiltonians in the presence of
superconducting gap. It was found that the bound state energies
follows the linear relationship with the angular momentum for both
cases when the chemical potential is highly positive, $\mu \gg
\Delta_0$. If the chemical potential is highly negative, $\mu \ll
-\Delta_0$, the linear relationship holds for the Dirac
Hamiltonian due to the particle-hole symmetry; whereas the bound
states for the Schrodinger Hamiltonian merge into the continuum.
After applying the external magnetic field perpendicular to the
interface, the bound states energies for the Dirac Hamiltonian are
insensitive; whereas there is a Zeeman shift for the Schrodinger
Hamiltonian. The bound states should be determined by the local
density of states on the surface from the scanning tunneling
microscopy experiments. The bound state energy differences can be
determined by the absorption of circular polarized light. The
qualitative differences between the Schrodinger and Dirac
Hamiltonian could serve as an indirect evidence of the existence
of Majorana fermions.

\section{acknowledgement}
I would thank very much Sungkit Yip for his help to my
understanding of the subject matter, and further discussion. The
work was supported by the National Science Council of the Republic
of China.

\vspace{15pt}

\vspace{15pt}

\vspace{15pt}

\vspace{15pt}

\vspace{15pt}

\vspace{15pt}

\vspace{15pt}

\vspace{15pt}

\vspace{15pt}

\end{document}